\documentclass[twocolumn,amsmath,amssymb,]{revtex4-2}

\usepackage{graphicx}
\usepackage{caption}
\usepackage{subcaption}

\usepackage{dcolumn}
\usepackage{bm}

\begin{document}

\title{Minimal Acquisition Time Polarized Neutron Imaging of Current Induced Magnetic Fields in Superconducting Multifilamentary YBCO Tape}


\author{Cédric Holme Qvistgaard}
\affiliation{DTU Energy, Technical University of Denmark, Kgs. Lyngby, Denmark}%

\author{Luise Theil Kuhn}
\affiliation{DTU Energy, Technical University of Denmark, Kgs. Lyngby, Denmark}

\author{Morten Sales}
\affiliation{
DTU Physics, Technical University of Denmark, Kgs. Lyngby, Denmark
}%

\author{Takenao Shinohara}
\affiliation{%
 J-PARC Center, Japan Atomic Energy Agency (JAEA), 2-4 Shirakata, Tokai, Ibaraki 319-1195, Japan}%

\author{Anders C. Wulff}
\affiliation{SUBRA A/S, Bygmarken 4, 3520 Farum, Denmark}

\author{Mette Bybjerg Brock}
\affiliation{SUBRA A/S, Bygmarken 4, 3520 Farum, Denmark}

\author{Søren Schmidt}%
\affiliation{ESS ERIC, Lund, Sweden}

\date{\today}

\begin{abstract}
In this paper we showcase the strengths of polarized neutron imaging as a magnetic imaging technique through a case study on field-cooled multifilamentary YBCO tape carrying a transport current while containing a trapped magnetic field. The measurements were done at J-PARC's RADEN beamline, measuring a radiograph of a single polarization component, to showcase the analysis potential with minimal acquisition time. 

Regions of internal damage are easily and accurately identified as the technique probes the internal magnetic field of the sample, thereby avoiding surface-smearing effects. Quantitative measurements of the integrated field strength in various regions are  acquired using time-of-flight information. Finally, we estimate the strength of the screening currents in the superconductor during the experiment by simulating an experiment with a model sample and comparing it to the experimental data. With this, we show that polarized neutron imaging is not only a useful tool for investigating magnetic structures but also for investigating samples carrying currents.

\end{abstract}

\maketitle


\section{Introduction}
Due to their magnetic moment, neutrons will interact with any magnetic field they pass through, allowing them to act as a direct probe for magnetic field strength. This forms the basis for the technique known as Polarized Neutron Imaging (PNI), used to image magnetic fields \cite{Kardjilov2008}.

When interacting with a magnetic field non-parallel to the direction of the neutron's spin, the magnetic moment of the neutron will start Larmor-precession in the external field, precessing in the plane perpendicular to the field direction. The precession angle follows the relation 
\begin{equation}
    \theta =  \gamma tB
    \label{Eq:simple_theta},
\end{equation}
with $\gamma$ being the gyromagnetic ratio, $t$ being the duration spent in the magnetic field, and $B$ the strength of the field. By polarizing a neutron beam such that all the magnetic moments align, one can probe the entirety of the magnetic field by measuring the degree by which the magnetic moments have rotated. 

In practice, the polarization of a neutron beam is measured indirectly by filtering a single spin component at a time. The fraction of a beam with incoming polarization direction \textit{i}, that aligns with the filter direction \textit{j} after sample interaction, is called the polarization degree of the beam $P_{ij}$. When $P_{ij}= $ 1 it means that the magnetic moment of the entire beam is parallel with the filter direction, for $P_{ij}= $ 0 it is  perpendicular to the field direction; for $P_{ij}= $ -1 it is anti-parallel.

Viewing the precession in terms of spherical coordinates, we know that the moment, and thus polarization, is a unit vector, and assuming \textit{i} to be the incoming polarization axis in a right-handed Cartesian coordinate system, we can define $\theta$ as the precession angle in the \textit{ij} plane, and $\phi$ as the precession angle in the \textit{jk} plane. With that, we can tie the polarization degree to the precession with the following relations, 
\begin{align}
\label{eq:FullP1}
&P_{ii} = \mathrm{cos}(\theta) \\
\label{eq:FullP2}
&P_{ij} = \mathrm{sin}(\theta)\mathrm{cos}(\phi) \\ 
\label{eq:FullP3}
&P_{ik} = \mathrm{sin}(\theta) \mathrm{sin}(\phi)
\end{align}

Assuming relatively weak and simple magnetic fields, such that we can ignore cascading effects resulting from the precession itself, one can write a simplified expression for the moment precession as

\begin{equation}
    \theta_i = \frac{\gamma m}{h} \lambda \int{(B_j + B_k) ds},
    \label{Eq:Theta}
\end{equation}

where $h$ is Planck's constant, $m$ the neutron mass and $\lambda$ the neutron wavelength. $\int{(B_j + B_k) ds}$ is the integrated field strength of the \textit{j} and \textit{k}-component of the magnetic field along the neutron flight path. The strength relation between $B_j$ and $B_k$ dictates the $\phi$ angle, meaning that if $B_k$ is dominant only $P_{ij}$ would be measurable, whereas a dominant $B_j$ component would show up in the $P_{ik}$ measurement.

Using this technique one can retrieve spatial information about the magnetic field along the neutron flight path, allowing for direct probing of the magnetic field inside a sample. It has been shown that a full 3D reconstruction is possible using polarized neutron tomography \cite{Sales2018} \cite{Busi2023}. Acquiring a full tomography is, however, a time-consuming process, in principle requiring 18 measurements for each tomographic angle; a spin up and down filtration to get each of the 9 polarization combinations of the polarization matrix. The focus of this paper will be showcasing the power of polarized neutron imaging measuring only a single polarization component from a single angle, thereby greatly reducing the acquisition time, though also reducing the spatial information gained.

To do this a multifilamentary YBCO superconducting tape described in great detail in \cite{Wulff2021} was chosen. This type of sample has already been studied extensively with other magnetic measurement techniques such as magneto-optics and vibrating sample magnetometry.\cite{Insinga2019}. This is advantageous as we can then model the sample structure to be applied in the analysis, and we can contextualize the results from the analysis by comparing them with the expected values from the literature. Secondly, the sample also lends itself well to polarized neutron radiography, as the planar geometry limits the possible position of the currents in space, making the source of the magnetic field well-known while still preserving interesting field dynamics inside the sample that other probing techniques cannot easily access. Finally, the superconductor allows us to investigate two different types of magnetic fields: The trapped magnetic field generated by screening currents in the superconductor, at low temperatures, as well as the induced magnetic field that occurs when a transport current is applied through the superconductor. 

\section{Materials and Methods}

The experiment was conducted at J-PARC's RADEN beamline using the polarimetric neutron imaging setup \cite{Shinohara2017}. The neutron beam was a pulsed beam and the wavelength was derived by Time-of-Flight (ToF) analysis. For this experiment it was set to span wavelengths from 1.5\,Å - 7.1\,Å split among 236 measurement bins. The sample consisted of a 20\,mm x 10\,mm x 91.2\,µm piece of multifilamentary YBCO, wherein the superconducting YBCO layer was only 1.2µm thick. The tape consists of 13 separate 0.8\,mm wide YBCO filaments each electrically isolated from one another and alternating in height, shown in the sketch in figure \ref{fig:sketch}. In a 1\,T field the tape was estimated to have a critical current of 650\,A at 4.2\,K, corresponding to a critical current of 50\,A per filament \cite{Insinga2019}. Figure \ref{fig:sample} shows a photo of the mounted sample. Two wires were soldered to the upper 3 filaments of the sample allowing the application of a transport current. The sample was placed perpendicular to the neutron beam (Z-direction) in what we will define as the "XY" plane, with X representing the horizontal direction and Y the vertical. 

The neutron detector used was a micro-channel plate (MCP) detector comprised of 512 x 512 pixels with a pixel size of 55\,µm x 55\,µm and a field-of-view of 28.2\,mm x 28.2\,mm \cite{Tremsin2009}. The imaging beamline was set to have pinhole-sample distance L=18\,m, and a pinhole size D=34.5\,mm giving an L/D ratio of 520 with the sample to detector distance being 0.5\,m. The spin up and down open beam measurements for the YY polarization were acquired for 2700 seconds each.

A 1\,T magnetic field was applied to the sample while it was cooled from above its critical temperature $T_c$ down to 5\,K in a cryostat. The applied magnetic field was turned off and the remanent trapped field was imaged by PNI. Spin up and down measurements were acquired for the XY polarization for 1800 seconds each.

Then, a transport current of 1\,A was applied through the upper 3 filaments of the sample while maintaining the trapped field, and another set of measurements was acquired following the same procedure. Finally, the sample was heated above $T_c$ and subsequently cooled to 5\,K without any applied magnetic field, after which polarized neutron imaging of the sample containing only the transport current was done for the XY polarization with an extended acquisition time of 3600 seconds.

\section{Analysis}
\begin{figure}
\begin{subfigure}{.5\linewidth}
\centering
\includegraphics[width=0.78\linewidth]{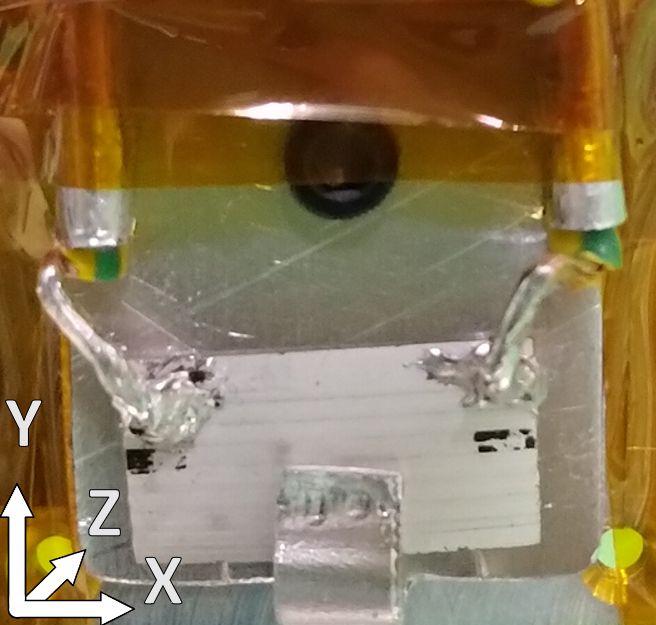}
\caption{}
\label{fig:sample}
\end{subfigure}%
\begin{subfigure}{.5\linewidth}
\centering
\includegraphics[width=1\linewidth]{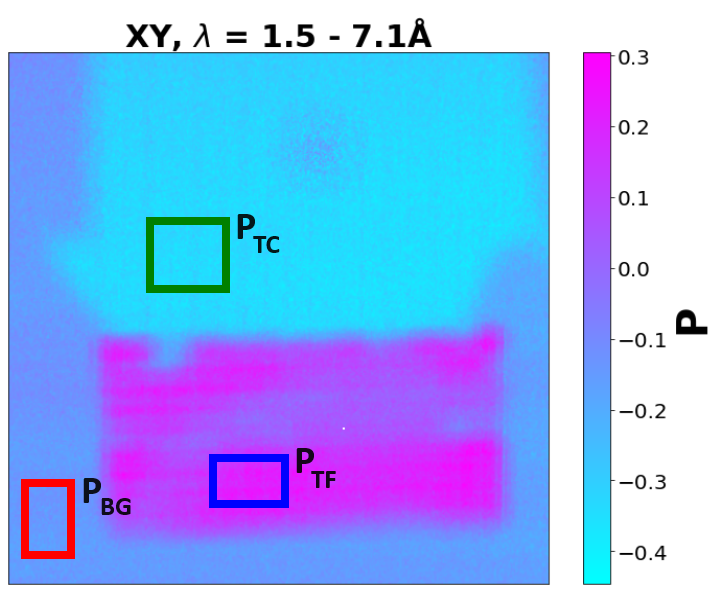}
\caption{}
\label{fig:data}
\end{subfigure}\\[1ex]
\begin{subfigure}{0.8\linewidth}
\centering
\includegraphics[width=1\linewidth]{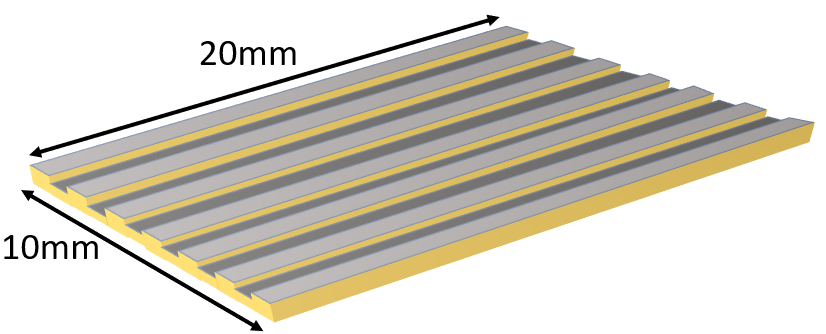}
\caption{}
\label{fig:sketch}
\end{subfigure}
\caption{\raggedright \textbf{a)} Photo of the mounted sample. \textbf{b)} XY polarization measurement for the full wavelength spectrum. Example regions of signal generated by Transport Current ($P_{TC}$), Trapped Field ($P_{TF}$) and Background ($P_{BG}$) are marked by a colored square. \newline \textbf{c)} Sketch of the sample consisting of 13 stripes on substrate separated in two height layers with marginal overlap.}
\label{fig:sample_and_data}
\end{figure}

Figure \ref{fig:sample} shows a photo of the sample, while \ref{fig:data} shows the XY polarization measurement of the sample containing both trapped magnetic field and transport current.

To calculate the polarization from the neutron radiographs we converted the imaging data using the following relation:
\begin{equation}
    P = \frac{\uparrow-\downarrow}{\uparrow +\downarrow}
\end{equation}
with the arrows indicating a measurement of the corresponding spin up($\uparrow$)/down($\downarrow$), and $P$ being the resulting polarization. For each of the 236 wavelength bins the polarization was normalized by dividing with the corresponding bin in the open beam polarization measurement.

In figure \ref{fig:data} we observe that the magnetic field generated by the transport current and the trapped magnetic field have opposite orientations, thereby making them easily distinguishable. The negative polarization signal in the top half of the image corresponds to the magnetic field induced by the transport current (TC) along the wires and upper 3 filaments of the sample, whereas the positive polarization signal at the bulk of the multifilament tape is caused by the trapped field (TF) inside the superconductor. Finally we also observe a weak background (BG) polarization signal in the areas surrounding the sample.

\subsection{Qualitative investigation}
Considering first the magnetic field generated by the transport current we notice that the measurements align well with the expectations: We observe a largely homogeneous signal in the region above the sample in between the two wire segments, which implies a homogeneous magnetic field along the Z-axis, as expected from three currents forming an U-shape, approaching the constant magnetic field generated by a current in a square loop.

Inspecting the part of the signal stemming from the trapped field, it is evident that the superconducting properties of the sample used were damaged. While the expectation was 13 identical stripes with a uniformly trapped magnetic field, we instead observed large patches of reduced signal in the middle regions of the YBCO tape and generally inhomogeneous polarization along each filament stripe. While the internal damaged region covers multiple filaments at the edges of the tape, the damage is concentrated to a few filaments, and when comparing with the photo of the sample in figure \ref{fig:sample} we observe that the damaged edge regions align with the black spots of the tape. These black spots are scratch damages to the silver (Ag) protection layer of the tape, likely caused by transportation. Thus we would expect hampered superconductive properties at these points. Here, the power of magnetic imaging is apparent, as we can conclude that the scratch damage is not limited to the scratched filaments but has cascading effects throughout the sample, degrading neighboring filaments that would otherwise look pristine. Thus PNI offers a direct probe of the screening current distribution, accurately visualizing the location of damage, as well as giving an indication of the degree of damage.

Finally, when inspecting the background region, we observe that a sizeable polarization signal of around 0.1 is present everywhere surrounding the sample. While it could be a residual signal from the trapped field and transport current, these two magnetic fields would have opposite orientations, and thus net different polarization measurements. As the background signal is homogenous throughout, and independent of applied current and trapped field, we must assume that it is a small systematic error overlaying all of the measurements; this will be discussed in detail below. 

\subsection{Time of flight analysis}
Going beyond the qualitative investigation, it is possible to retrieve the magnetic field strength by using equation (\ref{Eq:Theta}). It states that the resulting precession angle is linearly scaling with wavelength, with the proportionality being tied to the integrated magnetic field value along the neutron path $\int{B ds}$.
Utilizing the application of a pulsed neutron beam, which gives us the polarization degree as a function of wavelength, we can perform a linear regression and retrieve a quantitative measurement of the integrated magnetic field strength. 
To improve counting statistics, we have divided the ToF data into 11 bins covering 0.9\,Å each, utilizing only the 7 bins above 3.4\,Å with sufficient counting statistics for further analysis.

\begin{figure}
    \centering
    \includegraphics[width=0.95\linewidth]{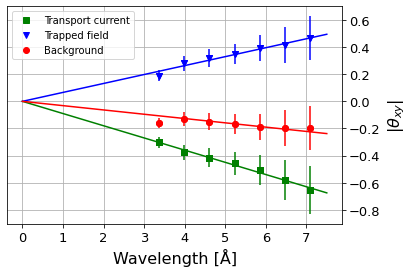}
    \caption{\raggedright Mean polarization of the three regions of interest shown in figure \ref{fig:data} at different wavelengths, alongside corresponding linear fit of the type $\theta = a\cdot \lambda$ with errorbars based on the standard deviation.
    }%
    \label{fig:TOF}
\end{figure}

To find the variation in time, a homogeneous section of the transport current (TC), trapped field (TF), and background (BG) was chosen, with the chosen sections marked by the colored boxes in figure \ref{fig:data}. The mean of each section was calculated to find their corresponding polarization ($P_{TC},P_{TF}, P_{BG})$, for each wavelength bin. The polarization was converted to a precession angle $\theta = \arcsin(P)$ and a linear regression of the type $\theta = a\cdot \lambda$ was then performed for each section, retrieving three fits. Finally, the slope value was divided by $\frac{\gamma m_n }{h}$ to retrieve $\int B_z ds$. In figure \ref{fig:TOF} $\theta_{xy}$ is plotted as a function of wavelength, alongside a fit extrapolated to $\lambda = 0$. In table \ref{tab:TOF} the fit parameters and resulting integrated magnetic field values for all three regions are shown alongside their statistical fit uncertainty.
\begin{table}[]
\begin{tabular}{|l|r|r|}
\hline
$\theta = a \cdot \lambda$ &$  a \,[\mathrm{rad/Å}] $& $   \int B_z \,ds \,[\mathrm{\mu T\cdot m}]$ \\
\hline
Transport Current (TC)               & -0.090(1)           & 1.94(2)                 \\ \hline
Trapped field (TF)             & 0.066(1)         & -1.42(3)               \\ \hline
Background (BG)             & -0.032(2)        & 0.68(4)               \\ \hline
TC corrected for BG       & -0.058(2)          & 1.26(5)                \\ \hline
TF corrected for BG       & 0.098(2)          & -2.11(3)              \\ \hline
\end{tabular}

\caption{\raggedright Linear fitting parameters, statistical uncertainties, and resulting integrated field strengths from the fit in figure \ref{fig:TOF} alongside fit parameters for the background corrected data.}
\label{tab:TOF}
\end{table}
We begin with analyzing the background. We observe that it is slowly increasing with wavelength, which clearly indicates the presence of a magnetic field outside the sample itself. As mentioned in the qualitative analysis, the homogeneity of the background field indicates that it is not a residual field from the sample. Furthermore, when inspecting the measurements without trapped field and without transport current, the background remains unchanged in the empty regions, proving that it is a consistent measurement error in the entire polarization image and in all measurements. To correct for this systematic error, we have subtracted the background for each wavelength bin. The resulting values with background correction are shown in table \ref{tab:TOF}.
The integrated magnetic field strength of the trapped field is 67\% stronger than the magnetic field generated by the transport current.

Given that the integrated magnetic field value from the 1\,A transport current is in the same order of magnitude as found in the superconductor, one might be tempted to conclude that the screening currents in the superconductor are far from their theoretical value of 50\,A. Such a conclusion would be faulty, exactly because field values provided in table \ref{tab:TOF} are field strengths integrated along the flight path, and not a direct field strength measurement at sample position. Due to the differing geometry of the current distributions, it is expected that the field from the large square loop of the transport current spatially spans a much larger area, than the strong but spatially concentrated field trapped inside the superconductor. 
In order to obtain a comparative analysis of the absolute magnetic field strengths, and an accurate estimate of the shielding current in the superconductor, further modelling and analysis is needed.

\subsection{Simulation of current distribution}
To be able to calculate the screening current strengths inside the superconducting filaments, we will in this section utilize modelling and simulations in conjunction with the measurements. 

Firstly, we utilize a theoretical model of the current flow inside a superconducting filament to simulate the resulting trapped magnetic field. Secondly, we perform a forward simulation of the experiment to compare the resulting polarization measurements of a certain current strength with those measured experimentally. Thus enabling an estimate of the value of screening current inside the filaments. 

\begin{figure}
    \centering
    \includegraphics[width=0.9\linewidth]{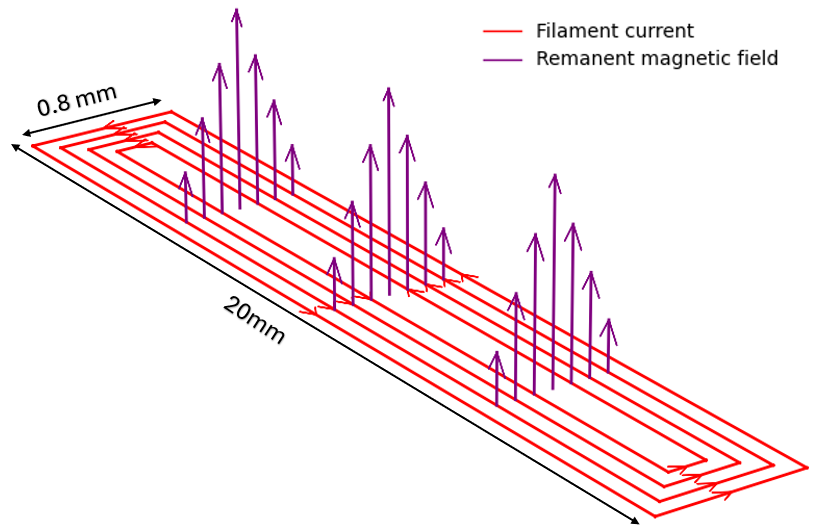}
    \caption{\raggedright Schematic configuration of screening current flow in a single model filament. Following the rooftop model, each filament consists of expanding square current loops parallel to the sides, all with the same current direction, emulating the effects caused by removing the applied field in a fully saturated regime.}
    \label{fig:Rooftop}
\end{figure}

To model the current flow in the sample we have used a combination of Bean's critical state model and the rooftop model with an example configuration shown in figure \ref{fig:Rooftop} \cite{Bean1964} \cite{Brandt1995}. For Bean's model, we have assumed that the applied field of 1\,T is much greater than the field needed to reach full field penetration in the sample, causing the entirety of the filament to reach the vortex state. Thus critical current flow is assumed to be maintained throughout the entire sample, even after the external field has been removed. With the rooftop model, we describe this critical current flow in the filaments as a series of outwards expanding square loops, with the current always being parallel to the sides, thereby creating a \textit{nearly} homogeneous magnetic field along the filament length, that is perpendicular to the surface of the sample as shown in figure \ref{fig:Rooftop}. 

Using the python Magpylib package \cite{Ortner2020}, we simulated the current distribution and resulting magnetic field based on the model for the current path, after which we applied a forward model of a polarized neutron imaging experiment to the simulated field, retrieving the simulated results of a measurement on an ideal sample. The forward model consists of a simple loop through each voxel in each $x,y$ row of the image, using the known neutron wavelength and voxel size to apply equation (\ref{Eq:simple_theta}) and update the magnetic moment accordingly. Running the simulation three times, initializing the simulation with the magnetic moment along $x,y,z$, then retrieves the entirety of the polarization matrix. This forward model emulates an idealized experiment and does not account for neutron beam fluctuations nor divergence. The geometric blur of the measurement is approximated by applying a Gaussian filter to the simulated magnetic field with $\sigma = 3.4$ and a truncation of three standard deviations. With the simulation being 300x300 pixels representing 28\,mm x 28\,mm these numbers result in a blur of 0.96\,mm, matching well the experimental blur of 0.96\,mm.
\begin{figure}
    \centering
    \includegraphics[width=1\linewidth]{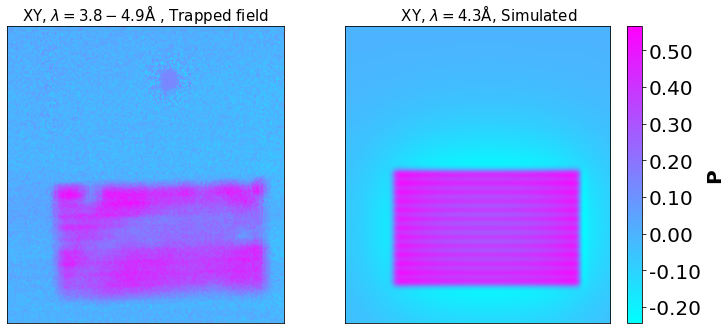}
    \caption{\raggedright Comparison between $P_{XY}$ for the background corrected trapped field measurement, and for and trapped field simulation at 5.4\,A per filament.}
    \label{fig:Simul}
\end{figure}

Tuning the simulated current strength until the simulated polarization agrees with the background corrected measurements of the exclusively trapped magnetic field, we can retrieve an estimate for the current strength generating the trapped field during the experiment. A comparison between the measurement and the simulation is shown in figure \ref{fig:Simul}, and using a total loop current of 5.4\,A pr filament, we retrieved a maximum XY-polarization of 0.56, which was equal to the largest experimental value measured. For the simulation, we retrieved an integrated field value of -2.28\,µTm which is close to the measured value of -2.11\,µTm, considering that the measurement was performed on a segment, and the simulation tries to match the highest pixel value, thereby overshooting a bit.  

The simulation agrees overall well with the experiments. Qualitatively we observe structural agreement between the simulated and experimental results, clearly showing the filament structure in both cases. With the highest estimated filament current being 5.4\,A, the estimate is almost an order of magnitude lower than the theoretical 50\,A filament capacity, even when accounting for the many uncertainties that reconstruction from modelling brings. 
This discrepancy makes sense when considering the damaged state of the sample, and simultaneously makes it evident that the sample damage has affected the entirety of the sample, and not only the central regions with lower signal, as one might otherwise conclude purely from the visual analysis.

\section{Discussion}

The first thing to note is that in the ToF analysis, we assumed that each region was magnetically independent; in reality, however, there could be a residual signal from the transport current overlaying the region of the trapped field and vice versa. However, we did also have access to measurements with purely trapped field or induced field from the transport currents, and when comparing the results in the regions of interest from these measurements with those found in the initial analysis, there were no significant differences in the results, indicating that each of the magnetic fields regions were well isolated from one another. 

Another concern one could have is omitting of the fitting constant $b$ in the application of a linear regression of the type $\theta = a\cdot \lambda + b$. This was done to add a physical constraint to the fit, as any exterior magnetic fields in the measurement should still tend towards zero for an infinitely fast neutron, and thus a constant offset on the fit would be nonphysical. Instead, we adjusted for the background signal by subtracting the signal from the background region. With this, we assume that the background field was the only systematic error in the measurement. A small mistiming in the spin flipper would add a constant offset $b$ in the time of flight profile of the neutron beam, however we have not observed such an error in the PNI setup, and we thus decided that removing the fitting constant was the best way to proceed with the analysis. 

When describing the constant background signal in the measurement, we accredited it to a constant exterior magnetic field, as the background polarization increased with wavelength, which would not happen if a miscalibration of the spin flipper had been at play. Considering what caused this background field, we assume that the majority can be accredited to the Earth's magnetic field and that the sample was mounted in a cryostat. Normally, in a PNI experiment, the sample and the region between the two spin analyzers are enclosed in a magnetically shielding $\mu$-metal box to prevent the effects of the Earth's magnetic field. However, in this experiment, to make space for the cryostat, a hole was cut in the upper and lower section of the shielding box, potentially allowing some residual magnetic field to enter the sample region and thus affect the measurements. 

It should be stressed that the estimates for the current strength in the filaments is ultimately a derived value from the simulation and model, and not a direct measurement. The accuracy of these estimates is fully dependent on the model and simulation used, and in this case, several simplifications were done.
For instance, to validate the simulations used to determine the current strength we also simulated a transport current of a similarly sized square loop with a current of 1\,A. Simulating this simple setup, the simulated integrated field strengths were about 30\% less than those measured in the experiment, indicating that there were experimental conditions that could not be fully replicated by the simulations, which should be taken into account for future work. The model of screening current was also simplified, affecting the accuracy of the estimates. We note in particular that the magnetic field in the sample is not homogeneous like the modelled one, thereby potentially deviating significantly, even along the least damaged parts. Furthermore, we also note that Bean's model assumes a thick slab of superconductor and not a thin tape. Finally, it disregards any additional effects that might be caused by surrounding magnetic fields. Thus much would be gained in terms of accuracy by using a more detailed experimental model.

Even more accurate results could have been achieved had a time-consuming full 3D polarization analysis been done on the sample. However, the aim of this work is not to model or analyze the microstructural details of the sample, but instead to illustrate the strengths of single projection polarized neutron imaging and its ability to yield fast quantitative estimates of current strengths using even relatively rudimentary data analysis and modelling.

\section{Conclusion}
In this study, we have showcased various strengths of the polarized neutron imaging technique alongside some of the most accessible analysis methods in the field. 
We demonstrated the qualitative investigation of sample quality through inspection of the polarized neutron image, observing significant regions of poor field retention in the superconducting tape. 
Furthermore, we retrieved a quantitative measure for the integrated strength of the trapped field inside the superconducting sample of $\mathrm{-2.11\,µTm}$, by utilizing time-of-flight analysis based on fast single projection polarized neutron imaging. 
Finally, we illustrated the necessity of using simulations in conjunction with the experimental data, allowing for the retrieval of derived experimental values that were not directly measured. By combining a theoretical model of the screening current structure in the sample with a simulated forward model of the experiment, the integrated screening current in each filament was estimated to an approximate value of of 5.4\,A.

\begin{acknowledgments}
We acknowledge support from the ESS Lighthouse on Hard Materials in 3D, SOLID, funded by the Danish Agency for Science and Higher Education, grant number 8144-00002B, and the support from the Danish Agency for Science, Technology, and Innovation for funding the instrument center DanScatt. We also acknowledge that the neutron experiment at the Materials and Life Science Experimental Facility of the J-PARC was performed under user program 2018L0502. We thank M. Strobl and A. Insinga for fruitful scientific discussions. 
\end{acknowledgments}

\bibliography{apssamp}

\end{document}